\documentclass[twocolumn,floatfix,prl,aps,showpacs]{revtex4}
\usepackage{graphicx,amsmath,amssymb}

\begin{document}

\title{Possible Anti-Pfaffian Pairing of Composite Fermions in the Lowest Landau Level}
\author{Sutirtha Mukherjee,$^1$ Sudhansu S. Mandal,$^1$ Arkadiusz W\'ojs,$^2$ and Jainendra K. Jain$^3$}
\affiliation{$^{1}$Department of Theoretical Physics, Indian Association for the Cultivation of Science, 
         Jadavpur, Kolkata 700 032, India}
\affiliation{
   $^{2}$Institute of Physics, 
   Wroclaw University of Technology,
   50-370 Wroclaw, Poland}
\affiliation{
   $^{3}$Department of Physics, 
   104 Davey Lab, 
   Pennsylvania State University, 
   University Park PA, 16802}
\date{\today}

\begin{abstract}
We predict that an incompressible fractional quantum Hall state is likely to form at $\nu=3/8$ as a result of a chiral p-wave pairing of fully spin polarized composite fermions carrying four quantized vortices, and that the pairing is of the Anti-Pfaffian kind. Possible experimental ramifications are discussed.
\end{abstract}


\maketitle

It is believed that a gapped paired state of composite fermions underlies the fractional quantum Hall effect (FQHE) at the half filled second Landau level (LL), and the chiral p-wave nature of the pairing implies vortices supporting Majorana modes \cite{Moore91,Greiter91,Nayak96,Read00,Scarola00}. We identify in this paper another candidate for similar physics, namely the state at filling factor 3/8 in the lowest Landau level. Our calculations below provide evidence that for fully polarized electrons, this state maps into an effective filling factor $\nu^*=3/2$ of composite fermions (CFs) carrying two vortices (denoted $^2$CFs), and that the composite fermions in the half filled second CF-LL (called $\Lambda$L) capture two additional vortices to turn into higher order composite fermions ($^4$CFs) and condense into a paired FQHE state. 

There are two topologically distinct candidates for the paired CF state: the Pfaffian (Pf) \cite{Moore91} and the Anti-Pfaffian (APf) \cite{Levin07,Lee07,Bishara09}. For electrons in the second LL ($\nu=5/2$), the two are exactly degenerate in the absence of LL mixing, due to an exact particle hole symmetry for a two-body interaction. A 3-body term induced by LL mixing, which breaks particle hole symmetry, will favor one of these states, and recent calculations have investigated which one is favored under realistic conditions \cite{Wojs10,Rezayi10}. Two candidates are obtained also at 3/8 by composite-fermionizing the Pf and APf at 3/2. While either would presumably occur for some model interactions (although the wave functions are sufficiently complex that no interaction can be constructed for which these are {\em exact} solutions), our calculations show that the Coulomb interaction selects the APf state at 3/8.  Interestingly, LL mixing is not necessary for discriminating between the Pf and the APf at 3/8, because there is no exact symmetry relating the two; the particle-hole symmetry for composite fermions in the second $\Lambda$L is only approximate, indicating that the Coulomb interaction between electrons induces a complex effective interaction between composite fermions that automatically contains two-, three- and higher body terms. 

The gap at 3/8, and the difference between the Pf and the APf, are governed by extremely small energy scales, and a 
theoretical resolution of these states requires a precise and reliable quantitative treatment of the interaction. We will employ the standard spherical geometry for our calculations, which considers $N$ electrons moving on the surface of a sphere, subject to a magnetic field emanating from a magnetic monopole at the center of the sphere. The strength of the monopole is denoted by $Q$, where $2Q$, an integer, is the number of flux quanta passing through the surface of the sphere. We will assume that the spin degree is frozen, and the magnetic field is high enough that LL mixing is suppressed. 
The filling factor is defined as $\nu = \lim_{N\to \infty}\frac{N}{2Q}$. Composite fermions \cite{Jain89} experience an effective flux $2Q^*$ given by $2Q^*= 2Q-2(N-1)$. At half filled second $\Lambda$L, the composite fermions satisfy $2Q^*+2=2N_2+\lambda$, where $N_2$ is the number of composite fermions in the second $\Lambda$L and $\lambda$ is an integer ``shift."
This leads to the following relations at $\nu=3/8$:
\begin{equation}
2Q = {8N+\lambda-10\over 3},\; 2Q^*= {2N+\lambda-4\over 3}, \; N_2={N-\lambda+1\over 3}
\nonumber
\end{equation}
We refer to $2Q$ given by the above relation as the ``Pf flux" for $\lambda=-3$ and the ``APf flux" for $\lambda=1$. (At this stage, the terms ``Pf flux"  and ``APf flux" should be taken merely as convenient labels, and not to mean that the actual states at these fluxes are represented by the Pf and the APf wave functions.)

\begin{figure}
\vspace{1cm}\includegraphics[width=7cm,angle=0]{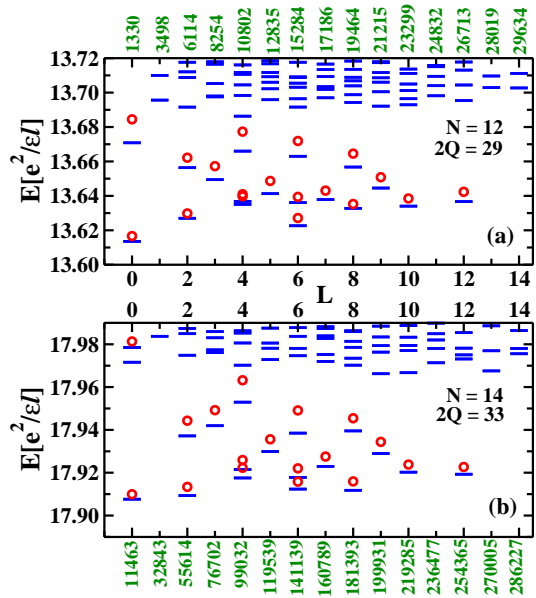}
\caption{Exact Coulomb spectra (dashes) at $[N,2Q]=$ [12,29] and [14,33], which correspond to APf and Pf fluxes at 3/8. Spectra obtained from composite fermions diagonalization are also shown (circles). The energies here and in Fig.~\ref{spectra} are the total Coulomb energies, which do not include the neutralizing background. The dimensions of the Hilbert space in the individual $L$ sectors are shown at the top and the bottom. \label{comp}}
\end{figure}

Exact diagonalization is possible for 14 electrons at Pf flux and 12 electrons at the APf flux, but not for larger systems \cite{Dim}. 
Further progress, however, can be made within the CF theory. We determine the energies and wave functions for low lying states by the method of CF diagonalization (CFD) \cite{Mandal02}, which proceeds along the following steps. We first perform exact diagonalization of the Coulomb Hamiltonian at $Q^*$ ($\nu^*=3/2$) keeping the lowest LL fully occupied, to obtain a basis \{$\Phi_{3/2}^{L,\alpha}$\}, where $\alpha$ labels the different basis functions in the total angular momentum $L$ sector. (Which interaction is chosen is unimportant because our goal is to produce all basis states with the lowest kinetic energy.) We then composite-fermionize this basis through the relation $\Psi_{3/8}^{L,\alpha} = P_{\rm LLL}\, \prod_{j<k}(u_jv_k-v_ju_k)^2\Phi_{3/2}^{L,\alpha}$, where $u=\cos\left(\theta/2\right)e^{-i\phi/2}$, $v=\sin\left(\theta/2\right)e^{i\phi/2}$, and $P_{\rm LLL}$ denotes projection of the wave function into the lowest LL, handled by the method in Ref.~\cite{JK}. (The $L$ quantum number is invariant under composite fermionization.) The correlated states \{$\Psi_{3/8}^{L,\alpha}$\} give us a basis for the low energy CF states at $\nu=3/8$.  All these states would be degenerate if composite fermions were non-interacting, but the degeneracy between them is split because of the residual interaction between composite fermions. We determine the low energy spectrum by diagonalizing the full Coulomb Hamiltonian in this CF basis. By construction, the basis functions are eigenstates of the total orbital angular momentum $L$, allowing us to diagonalize in each $L$ sector separately. The basis functions are very complex and non-orthogonal, but efficient methods have been developed for a Gram-Schmid orthogonalization and an evaluation of the Hamiltonian matrix by Metropolis Monte Carlo (which requires up to $10^8$ Monte Carlo steps for each system) \cite{Mandal02}. A diagonalization of this matrix produces the low energy spectra as well as eigenfunctions. These contain no adjustable parameters, and the Monte Carlo statistical uncertainty can be reduced to the desired level by increasing the number of iterations accordingly. We are able to study systems with as many as 26 particles, which allows us to draw what we believe to be reliable conclusions.

In Fig.~\ref{comp} we compare the CFD  spectra with those obtained from an exact diagonalization of the Coulomb interaction in the full lowest LL space for $N=14$ at the Pf flux and $N=12$ at APf flux. The following features can be noted. The exact spectrum contains a band of low energy states which has a complete one to one correspondence with the low energy band of non-interacting fermions at $Q^*$. The remarkable similarity between the low energy bands of the two spectra is thus nicely explained by the observation that, for non-interacting composite fermions these two would be related by an exact particle-hole symmetry for composite fermions in the second $\Lambda$L. The splitting of states in these bands, which arises due to inter-CF interactions, is not identical, however, indicating that the particle hole symmetry for composite fermions is only approximate. Importantly for our purposes, the CFD spectra accurately reproduce the exact spectra including the very fine differences between them, in spite of the large dimensions of the Fock spaces in the relevant sectors (shown in Fig.~\ref{comp}). These results show that: (i) the physics of the 3/8 state is indeed described in terms of composite fermions; and (ii) the CFD produces essentially exact results.

\begin{figure}
\vspace{1cm}\includegraphics[width=8cm,angle=0]{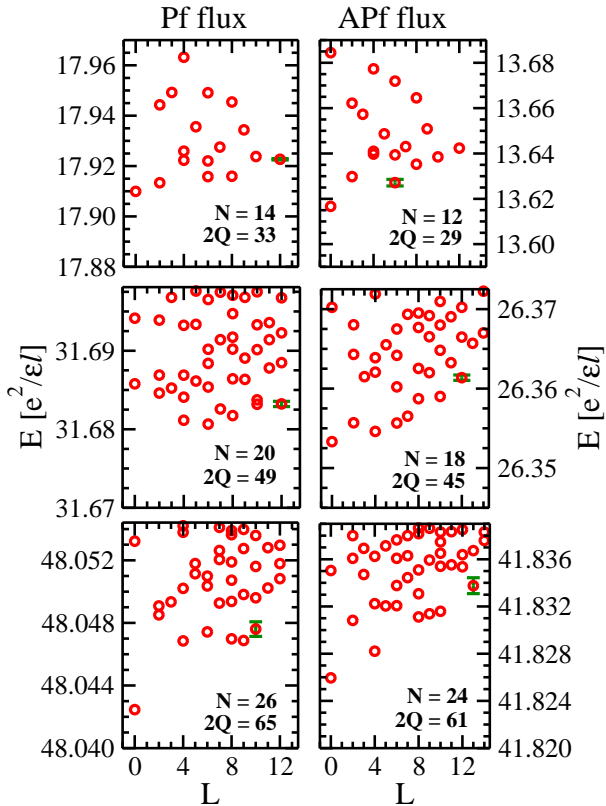}
\caption{Energy spectra obtained from CF diagonalization at both ``Pf flux" (left panels) and ``APf flux" (right panels) at $\nu=3/8$, with $N$ and $2Q$ values shown on the figure. To avoid clutter, the typical estimated statistical uncertainty from Metropolis Monte Carlo evaluation of integrals is shown only on one point. Only states below certain energy are shown.
\label{spectra}}
\end{figure}

A remark is in order regarding another method that has been used previously to treat interactions between composite fermions in a partially filled $\Lambda$L. In this method, the 2-body inter-CF interaction is determined \cite{Quinn96,Lee01,Wojs10} by considering two composite fermions in the relevant $\Lambda$L, and then making the assumption that the state of many composite fermions in that $\Lambda$L is governed by this interaction. This method provides a reasonably good first order approximation, but is less accurate than the one used here, and, in particular, cannot discriminate between the Pf and the APf states because it respects particle-hole symmetry for composite fermions.

CFD allows us to obtain the spectra for larger systems, shown in Fig.~\ref{spectra}. An incompressible state manifests itself through an $L=0$ (spatially uniform) ground state that is separated from others by a robust gap. The fact that all of the APf flux values produce incompressible states (but not all of the Pf flux values do) suggests that an incompressible state indeed occurs at 3/8 at the APf flux. The system sizes are still not large enough to be able to estimate the gap reliably, but we note that the gap to the lowest neutral excitation for the two largest systems is $\sim$0.002 $e^2/\epsilon l$, which we take as a measure of the energy scales associated with this state. This is a factor of 50 smaller than the ideal gap of the nearby 1/3 state, and a factor of 10 smaller than the gap at 5/2, thus indicating the fragile nature of the state.

For a further confirmation that the actual state is indeed described by the APf wave function, we construct the following trial wave functions, labeled 1 and 2, at the Pf and the APf flux values:
\begin{equation}
\Psi_{3/8}^{\rm trial-1}=P_{\rm LLL}\, \prod_{j<k}(u_jv_k-v_ju_k)^2\Phi_{\rm 3/2}^{\rm Pf/APf} \nonumber
\end{equation}
\begin{equation}
\Psi_{3/8}^{\rm trial-2}=P_{\rm LLL}\, \prod_{j<k}(u_jv_k-v_ju_k)^2\Phi_{\rm 3/2}^{\rm Coulomb} \nonumber
\end{equation}
Here, $\Phi_{\rm 3/2}^{\rm Pf/APf}$ is the Pf or APf wave function at 3/2, which refers to the state in which the lowest LL is fully occupied and the electrons in the second LL form a Pf or an APf state. (We produce the Pf state in the lowest LL by diagonalizing the 3-body interaction Hamiltonian \cite{Greiter91} $V_3=\sum_{i<j<k}P^{(3)}_{ijk}(3Q-3)$, where $P^{(3)}_{ijk}(L)$ projects the state of the three particles $(i,j,k)$ into the subspace of total orbital angular momentum $L$; the APf state is obtained by its particle hole conjugation; we then elevate the Pf/APf to the second LL and fill the lowest LL fully to obtain $\Phi_{\rm 3/2}^{\rm Pf/APf}$.) The wave function $\Phi_{\rm 3/2}^{\rm Coulomb}$ is the exact Coulomb eigenstate at the relevant $Q^*$ at $\nu^*=3/2$. Composite-fermionization of these wave functions gives two trial wave functions at 3/8. Tables I and II compare the energies of these trial wave functions with the CFD energies, and also give the overlaps of these trial wave functions with the CFD wave function.  The APf state has higher overlaps, again indicating that it is favored over the Pf. The overlaps are not extremely high, but are decent, roughly on the same order as the overlaps of the 5/2 Coulomb ground state with the Pf / APf wave function. Taking into account all of these facts, we conclude that it is likely that the 3/8 state is incompressible and described by a composite-fermionized APf state.

\begin{table}
\caption{Comparing the CFD ground state $\Psi_{3/8}^{\rm CFD}$ at at the ``Pf flux" $2Q=(8N-13)/3$, obtained by CF diagonalization, with the trial wave functions, $\Psi_{3/8}^{\rm trial-1}$ and $\Psi_{3/8}^{\rm trial-2}$, derived from the composite fermionization of the Pf and the exact Coulomb states at 3/2. (See text for definition.) $E_{3/8}^{\rm CFD}$, $E_{3/8}^{\rm trial-1}$, and $E_{3/8}^{\rm trial-2}$ are the energies per particle for these three states, quoted in units of $e^2/\epsilon l$, where $l=\sqrt{\hbar c/eB}$ is the magnetic length and $\epsilon$ is the dielectric constant of the background material; this energy includes the interaction with the positively charged background. The numbers $O_j=\langle \Psi_{3/8}^{\rm trial-j}|\Psi_{3/8}^{\rm CFD}\rangle$ are the overlaps of the two trial wave functions with the CFD ground state (all properly normalized). For $N=20$ the comparisons are given for the lowest energy state in the $L=0$ sector; the CFD ground state occurs at $L=6$. \label{tab:Pf}}
\begin{tabular}{| c | c | c | c | c | c |} \hline
$N$ & $O_1$ & $O_2$  & $E_{3/8}^{\rm trial-1}$ & $E_{3/8}^{\rm trial-2}$ & $E_{3/8}^{\rm CFD}$\\ \hline
14  & 0.726(1) & 0.973(2) & -0.44153(8)& -0.44372(9)& -0.44403(9) \\
20* & 0.379(1) & 0.434(1)  & -0.43418(2)& -0.43515(8)& -0.43599(1)\\
26  & 0.271(1) & 0.526(1) & -0.43021(9)& -0.43146(6)& -0.43248(4) \\ \hline
\end{tabular}
\end{table}

\begin{table}
\caption{Comparing the CFD state at ``APf flux" $2Q=(8N-9)/3$ with two trial wave functions, $\Psi_{3/8}^{\rm trial-1}$ and $\Psi_{3/8}^{\rm trial-2}$, obtained by composite fermionization of the APf and the exact Coulomb states at 3/2. Other symbols have the same meaning as in Table \ref{tab:Pf}. \label{tab:APf}}
\begin{tabular}{| c | c | c | c | c | c |}\hline
$N$ & $O_1$ & $O_2$  & $E_{3/8}^{\rm trial-1}$ & $E_{3/8}^{\rm trial-2}$& $E_{3/8}^{\rm CFD}$ \\ \hline
12  & 0.816(1) & 0.994(1)  & -0.43903(2)& -0.44076(6)& -0.44079(9) \\
18  & 0.587(2) & 0.622(2) & -0.43168(9)& -0.43225(7) & -0.43310(8) \\
24  & 0.503(1) & 0.781(1) & -0.42845(9)& -0.42948(8) & -0.42995(7) \\ \hline
\end{tabular}
\end{table}

The principal consequences arising from our calculations above are that (i) FQHE is possible at 3/8; (ii) it originates due to p-wave pairing of composite fermions in the second $\Lambda$ level; and (iii) the pairing is of the APf type. We now discuss how these can be tested by experiment. (i) Pan {\em et al.} observed \cite{Pan03} a resistance minimum at 3/8 in 2003, but a well quantized plateau has not been seen so far. Given the rather small gap, 3/8 FQHE would require pristine conditions, and we hope that the interesting physics of this fraction will stimulate further investigation. The FQHE should occur at sufficiently large magnetic fields where the state is fully spin polarized; the spin polarization of the state can be measured by NMR or by optical means, as for 5/2 \cite{52spin}. (ii) The chiral p-wave pairing reflects through the charge and non-Abelian braid statistics of the quasiparticles. The excess charge associated with an excitation can be seen to be $e/16$ by one of many methods. The non-Abelian braid statistics of the excitations will have similar signatures as those predicted for 5/2 \cite{SDS05,Stern07}.  (iii) The above features do not distinguish between whether the state is Pf or APf. Proposals have been made on how the Pf and the APf states at 5/2 can be distinguished experimentally through their different edge structures \cite{Wen93,Levin07,Lee07,Wen92}, and these analyses carry over to the 3/8 state with appropriate modifications. The Pf and APf states at 3/2 have edge structures (disregarding the possibility of edge reconstruction) 3/2(Pf)-1-0 and 3/2(APf)-2-1-0, respectively, which translate, upon composite-fermionization, into 3/8(Pf)-1/3-0 and 3/8(APf)-2/5-1/3-0 at 3/8. An immediate consequence is that the APf will {\em necessarily} contain counter-propagating edge modes, including an up-stream charge neutral Majorana mode, which can have experimental signatures, e.g., in noise measurements in an upstream voltage contact \cite{Bid}. Observation of such modes would not constitute a proof of APf, because the Pf state can also have backward moving modes due to edge reconstruction. However, we expect that the physics of edge reconstruction at 3/8 should not be too different from that at the nearby fractions 1/3 or 2/5, so an observation of counter-propagating modes at 3/8 concurrent with an absence of such modes at 1/3 and 2/5 can be taken as a substantial evidence for APf state at 3/8. The thermal Hall conductivity $K_H=\partial J_Q/\partial T$, where $J_Q$ is the thermal energy current and $\partial T$ is the ``Hall" temperature difference, can also in principle distinguish between the Pf and the APf  \cite{Levin07}. In units of $(\pi^2 k_B^2 /3h)T$, each chiral boson edge mode contributes one unit and the Majorana fermion mode 1/2 unit \cite{Kane,Kitaev}, with the sign depending on the direction of propagation.  The boundary 3/8(Pf)-1/3 supports a chiral boson and a Majorana mode; the boundary 3/8(APf)-2/5 also supports a chiral boson and a Majorana mode, but moving in the upstream direction. This produces thermal Hall conductivity of $1+1/2+1=5/2$ for the Pf and $-1-1/2+1+1=1/2$ for the APf at 3/8. This result is believed to be robust against interactions, disorder or edge reconstruction. One may also consider various tunneling exponents, following Wen \cite{Wen92,Wen93}. The exponent describing the long distance decay of the propagator of the charge 1/16 nonabelian quasiparticles can be shown \cite{exponent} to be $g=7/13$ for the 3/8 Pf; this exponent appears in the prediction \cite{Wen92}, assuming absence of edge reconstruction, that the current from one edge of the sample to the opposite edge near a quantum point contact satisfies $I\sim V^{2g-1}$ and the tunnel conductance has a temperature dependence $\sigma\sim T^{2g-2}$. For the APf state, on the other hand, the presence of up-stream neutral modes renders the various exponents non-universal even for an unreconstructed edge.

There have been previous studies of 3/8 within a model that treats the effective interaction between composite fermions through a two-body term. Ref.~\cite{Lee01} found, by comparing several variational wave functions, that for a fully polarized state the stripe phase has lower energy than the Pfaffian. Using the same model interaction, Ref.~\cite{Wojs04} studied the 3/8 sate, but the model does not distinguish between Pf and APf and also does not produce incompressible states at all even $N$. While we believe that our current treatment is more reliable for reasons mentioned above, we obviously cannot rule out that the system sizes considered here may not capture the true nature of the thermodynamic phase, and the ultimate resolution to this issue will likely come from experiments. An earlier study \cite{Scarola02} considered composite fermions in the spin reversed $n=0$ $\Lambda$L, also using a 2-body interaction model for composite fermions, and pointed toward a partially spin polarized paired FQHE state (without distinguishing between Pf or APf); that physics is not relevant at very high magnetic fields where the electrons are fully spin polarized. 

Before closing, we note that we have not included in our work the effect of finite thickness, LL mixing, and disorder. While these will surely make quantitative corrections, we do not see any reason why they should change the qualitative physics of the state.  

We thank Michael Levin for a valuable discussion. We acknowledge financial support from the NSF under grant no.
DMR-1005536, the DOE under Grant No. DE-SC0005042, and the Polish NCN grant 2011/01/B/ST3/04504. The computation was performed using the cluster of the Department of Theoretical Physics, Indian Association for the Cultivation of Science and the Wroclaw Centre for Networking and Supercomputing.


\begin{thebibliography}{99}

\bibitem{Moore91} G. Moore and N. Read, Nucl. Phys. B {\bf 360}, 362 (1991).

\bibitem{Greiter91} M. Greiter, X.-G. Wen, and F. Wilczek, Phys. Rev. Lett. {\bf 66}, 3205 (1991).

\bibitem{Nayak96} C. Nayak and F. Wilczek, Nucl. Phys. B {\bf 479}, 529 (1996).

\bibitem{Read00} N. Read and D. Green, Phys. Rev. B {\bf 61}, 10267 (2000).

\bibitem{Scarola00} V. W. Scarola, K. Park, and J. K. Jain, Nature {\bf 406}, 863 (2000).

\bibitem{Levin07} M. Levin, B. I. Halperin, and B. Rosenow, Phys. Rev. Lett. {\bf 99}, 236806 (2007).

\bibitem{Lee07} S.-S. Lee, S. Ryu, C. Nayak and M. P. A. Fisher, Phys. Rev. Lett. {\bf 99}, 236807 (2007).

\bibitem{Bishara09} W. Bishara and C. Nayak, Phys. Rev. B {\bf 80}, 121302(R) (2009).

\bibitem{Wojs10} A. W\'ojs, C. T\"oke and J. K. Jain, Phys. Rev. Lett. {\bf 105}, 096802 (2010).

\bibitem{Rezayi10} E. H. Rezayi and S. H. Simon, Phys. Rev. Lett. {\bf 106}, 116801 (2011).

\bibitem{Jain89} J. K. Jain, Phys. Rev. Lett. {\bf 63}, 199 (1989).

\bibitem{Dim} This can be seen by considering the next available systems. For $N=18$ at $2Q=45$ (APf) the dimension of the Hilbert space in the $L_z=0$ subspace is greater than 25 billion, and splits into sectors counting $>$6 million states for $L=0$, $>$18 million states for $L=1$, etc. 
%6,090,215 to 178,078,145 between $L=0=L_z$ and $L=15,L_z=0$ sectors. 
For $N=20$ at $2Q=49$ (Pf) the $L_z=0$ subspace has dimensions greater than 368 billion, with 
%the $L=L_z=0$ sector containing 69,602,057 
$>$69 million independent states in the $L=0$ sector, $>$208 million states in the $L=1$ sector, and so on. These and larger systems are clearly out of the reach of exact diagonalization.

\bibitem{Mandal02} S. S. Mandal and J. K. Jain, Phys. Rev. B {\bf 66}, 155302 (2002).

\bibitem{JK} J.K. Jain and R.K. Kamilla, Phys. Rev. B. {\bf 55}, R4985 (1997); Int. J. Mod. Phys. B {\bf 11}, 
2621 (1997). 

\bibitem{Quinn96} P. Sitko, S.N. Yi, K.S. Yi and J.J. Quinn, Phys. Rev. Lett. {\bf 76}, 3396 (1996).

\bibitem{Lee01} S.-Y. Lee, V.W. Scarola, and J.K. Jain, Phys. Rev. Lett. {\bf 87}, 256803 (2001); Phys. 
Rev. B {\bf 66}, 085336 (2002). 

\bibitem{Pan03} W. Pan, H. L. Stormer, D.C. Tsui, L. N. Pfeiffer, K. W. Baldwin, and K. W. West, Phys. Rev. Lett. {\bf 90}, 016801 (2003).

\bibitem{52spin} L. Tiemann, G. Gamez, N. Kumada, and K. Muraki, Science {\bf 355}, 828 (2012); M. Stern, B. A. Pior, Y. Vardi, V. Umansky, P. Plochocka, D. K. Maude, and I. Bar-Joseph, Phys. Rev. Lett. {\bf 108}, 066810 (2012); U. Wurstbauer, K. W. West, L.N. Pfeiffer and A. Pinczuk, arXiv:1204.4759v1 (2012).

\bibitem{SDS05} S. Das Sarma, M. Freedman, and C. Nayak, Phys. Rev. Lett. {\bf 94}, 166802 (2005).

\bibitem{Stern07} A. Stern and B. I. Halperin, Phys. Rev. Lett. {\bf 96}, 016802 (2006)

\bibitem{Wen93} X. G. Wen, Phys. Rev. Lett. {\bf 70}, 355 (1993).

\bibitem{Wen92} X. G. Wen, Int. J. Mod. Phys. B {\bf 6}, 1711 (1992).

\bibitem{Bid} A. Bid, N. Ofek, M. Heiblum, V. Umansky, and D. Mahalu, Phys. Rev. Lett. {\bf 103}, 236802 (2009); A. Bid, N. Ofek, H. Inoue, M. Heiblum, C. L. Kane, V. Umansky, and D. Mahalu, Nature {\bf 466}, 585 (2010).

\bibitem{Kane} C. L. Kane and M. P. A. Fisher, Phys. Rev. B {\bf 55}, 15832 (1997).

\bibitem{Kitaev} A. Kitaev, Annals of Phys. {\bf 321}, 2 (2006).

\bibitem{exponent} Generalizing Wen's treatment of the 1/2 Pfaffian state \cite{Wen92,Wen93}, the edge of the 3/8 Pf is effectively described by ${\cal L}={\cal L}_1+{\cal L}_2$ containing two bosonic ($\phi_1$ and $\phi_2$) and one majorana ($\psi$) modes, with $4\pi {\cal L}_1= \sum_{i,j=1,2} \partial_{t}\phi_i K_{ij} \partial_x \phi_j - \partial_x \phi_i V_{ij} \partial_x \phi_j$ and ${\cal L}_2=i\psi (\partial_t-v\partial_x)\psi$. Here the $V$ is a 2$\times$2 interaction matrix, and $K$ is a 2$\times$ 2 matrix with diagonal elements 3 and 5, and both off diagonal elements 2. Using standard methods from conformal field theory, one can show that the nonabelian quasiparticle $\sigma e^{i\phi_1/2}$ ($\sigma$ is an operator that connects different sectors of $\psi$) has charge 1/16 and its propagator decays at long distances with exponent $g=7/32$.

\bibitem{Wojs04} A. W\'ojs, K.-S. Yi and J.J. Quinn, Phys. Rev. B {\bf 69}, 205322 (2004).

\bibitem{Scarola02} V. W. Scarola, J. K. Jain and E.H. Rezayi, Phys. Rev. Lett. {\bf 88}, 216804 (2002).


\end{thebibliography}
\end{document}